\newcommand{\vy}[2]{#1_{\scriptscriptstyle #2}}

\documentclass[12pt,preprint]{aastex}

\shorttitle{Ionization Driven Fragmentation of FeLoBAL Quasars}    
\shortauthors{Bautista \& Dunn} 
\begin{document}
\title{Ionization Driven Fragmentation of Gas Outflows Responsible for
FeLoBALs in Quasars} 

\author{Manuel A.\ Bautista$^{1,2}$ and Jay P. Dunn$^2$}
\affil{$^1$Department of Physics, Western Michigan University, Kalamazoo, MI 49008-5252\\
$^2$Department of Physics, Virginia Polytechnic and State University,
Blacksburg, VA 24061}

\begin{abstract}
We show that time variations in the UV ionizing continuum of quasars, on 
scales of $\sim$1~year, affect the dynamic structure of
the plasmas responsible for 
low ionization broad absorption lines. Variations
of the ionizing continuum 
produce non-equilibrium photoionization conditions over a significant fraction
of the absorbing clouds and supersonically moving ionization fronts.
When the flux drops the contraction of the ionized region drives a supersonic
cooling front towards the radiation source and a rarefaction wave in the opposite direction. The pressure imbalance is compensated by an increased speed of the
cool gas relative to the front. When the flux recovers the cool gas is re-ionized and re-heated by a supersonic ionization front traveling away from the radiation source and a forward shock is created. 
The reheated clouds equilibrate to a temperature of $\sim 10^4$~K
and are observed to have different radial velocities than the main cloud. 
Such fragmentation seems consistent with the multicomponent structure
of troughs seen in some objects. The velocity differences
measured 
among various components in the quasars QSO 2359--1241
and SDSS J0318--0600 can be reproduced by our model
if strong magnetic fields ($\sim$10~mG) are present within the clouds.
\end{abstract}

\section{Introduction}

Gas outflows 
a potential source of feedback that regulates the evolution of the central 
black hole, the host galaxy, and the intergalactic 
medium (e.g. Elvis 2006). 
In the rest-frame UV spectra of 10 -- 20\% of all quasars \citep{knigge,
foltz},      
outflows are manifested in blueshifted Broad Absorption Lines (BAL). The most common of these
are associated with resonance lines of medium
ionization species, such as \ion{C}{4}, \ion{N}{5}, \ion{O}{6}, \ion{Si}{4}, and
\ion{H}{1} and 
can reach velocities as high as 50,000 km $s^{-1}$ \citep{weymann,
turnshek}. Unfortunately, their saturated nature limits their utility in
determining the physical characteristics of the outflow. 
A subset of BAL quasars, known as LoBALs, also show
complexes of narrower
(on the order of $\sim100$~km~s$^{-1}$)
absorption features from low ionization species
such as \ion{Mg}{2}, \ion{Al}{2}, and \ion{Si}{2}. 
Among these, objects that include 
troughs of metastable \ion{Fe}{2} or \ion{Fe}{3}  are specifically 
called FeLoBALs. These are useful 
because the column densities of metastable levels 
offer valuable diagnostics of the outflow 
(Gabel et al. 2006; Arav et al. 2005; and references therein). 
These studies place the gas responsible for the FeLoBAL at distances
between $\sim$ 1~kpc \citep{korista08, moe09} and up to $\sim$10 kpc 
\citep{dunn09}. 

We studied the various kinematic components in
the FeLoBAL troughs 
of two quasars, QSO 2359--1241
and SDSS J0318--0600, $z=$0.868 and 1.967 respectively \citep{bautista10}. 
It was found that  among the various kinematic components the highest 
column density component of each system also had
the highest particle density, while all other components have densities
of about one quarter 
of that of the main component. Moreover, all the components 
seem to be located at the same distance from the central source,      
suggesting that they are all related to each other.
This is also consistent with the finding of \cite{voit} that
BALs and LoBALs could be explained simultaneously by ablation of either
a single large cloud or many scattered clouds. 

Another characteristic of FeLoBALs is that, within the sample 
of known objects, they are borderline density-bounded.
FeLoBAL quasars are seen in absorption from 
species that arise from the ionization front (IF) of the cloud (e.g. \ion{Si}{2} and \ion{Fe}{2})
while atoms from a neutral cold region, e.g. \ion{C}{1} or \ion{Fe}{1}, are not 
seen. This seems to point out some relationship between the ionization front 
and the nature of the clouds.  
The location of the IF of a cloud of a given density depends on the flux of hydrogen ionizing photons, particularly in the $\sim$1 to 4 Ryd energy range, from the radiation source. 
However, quasar continua are known to vary in 
the optical and UV within time scales 
of the order of one year \citep{kaspi}.
Hence, it is important to understand the effect of the varying radiation field on the
structure of the FeLoBAL.

In this letter we show that when the flux that ionizes a FeLoBAL cloud drops the
depth of the ionized region often shrinks at supersonic speeds followed by
a cooling front, as explained in Sections 2 and 3. In section 4 we show that
the supersonic cooling front creates a rarefaction wave, traveling in the opposite
direction to the front, which results in fragmentation of the cloud. Under some 
conditions, such a fragmentation can explain the observed structure of FeLoBALs.

\section{Magnitude scales and the effect of flux variations on FeLoBALs}

\begin{figure}
\rotatebox{-0}{\resizebox{\hsize}{\hsize}
{\plotone{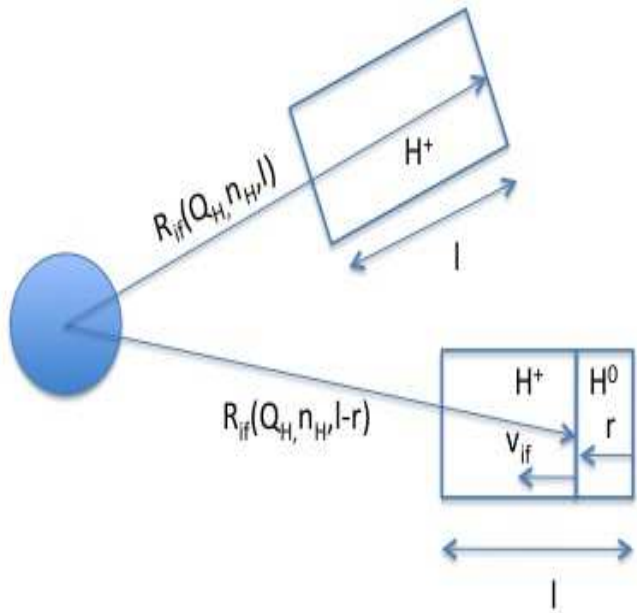}}}
\caption{Illustration of a photoionized cloud of depth $l$ traveling
away from the light source.
The upper cloud is right at $R_{if}$ and the
whole depth $l$ is ionized. In the lower  cloud only a depth $l-r$ is
ionized due to either a larger physical distance or a drop in
$Q_H$. 
}
\label{sketch}
\end{figure}

Let us start with the simple picture illustrated in Fig. \ref{sketch}. Here, $r$ is the depth of the neutral region of the cloud
and it is measured from the far end of the cloud towards the central source. 
For a given luminosity of the
source and a cloud of constant particle density $\vy{n}{H}$ moving in a medium of negligible density one can draw an imaginary radius
within which the cloud is fully ionized. This is
\begin{equation}
R_{if}(\vy{Q}{H},\vy{n}{H},l)= \left({\vy{Q}{H}\over 4\pi \vy{n}{H}^2{\alpha} l}\right)^{1/2},
\end{equation}
where $\vy{Q}{H}$ is rate of hydrogen ionizing photons from the central source, 
${\alpha}$ is the recombination rate coefficient, 
and $l$ is the depth of the cloud.
For a cloud located beyond $R_{if}$ the ionized depth ($l-r$) is related to
the physical distance ($R$) as 
$R^2(l-r)= \vy{Q}{H}/(4\pi \vy{n}{H}^2\alpha)$. 
Thus, FeLoBAL outflows that result from the formation of an IF, but 
without an extended neutral region, must be located right at 
$R_{if}$.
Then, a drop $\Delta\vy{Q}{H}$ in the ionization flux, 
i.e. $\vy{Q}{H}'=\vy{Q}{H}-\Delta\vy{Q}{H}$, will create a region of depth 
${r} = l\times {\Delta\vy{Q}{H}/\vy{Q}{H}}$ of either neutral or recombining plasma. 
Here, $l\approx {\vy{N}{H}/\vy{n}{H}}$, 
where $\vy{N}{H}$ is total column density of the cloud and has a typical value of $\sim 10^{20.5}$~cm$^{-2}$. 
In deriving this equation we assumed that $l<<R_{if}$ and $r<<l$. In
QSO 2359--1241 and SDSS J0318--0600 $\vy{n}{H}$ are $\sim 10^{4.4}$~cm$^{-3}$ and
$\sim 10^{3.4}$~cm$^{-3}$ and the sizes of the absorbers are $10^{16}$--$10^{17}$~cm (i.e. 
$\sim$0.01 to 0.1 light-years).
The change in the ionization of the gas is not instantaneous, but delayed 
by the recombination time scale of the plasma 
$\tau_{\alpha}=1/\vy{n}{H}\alpha$, with 
$\alpha$$\approx 2\times 10^{-13}$~s$^{-1}$cm$^{-3}$ 
at $T_e\approx 10^4$~K. 
Then, $\tau_{\alpha}\approx 10$~yrs in QSO 2359--1241 and 100~yrs in
SDSS J0318--0600.
For example, in the case that 
$\vy{Q}{H}$ is abruptly shut off, the size of the ionized region is
\begin{equation}
l-r(t)=l e^{-t/\tau_{\alpha}},
\end{equation}
where $t$ is the time measured from the moment that the flux is turned off. 
Thus, the depth of the ionized region shrinks with a speed 
\begin{equation}
v_{if}={d(l-r(t))\over dt}\approx {l\over \tau_{\alpha}}\ for\ t<<\tau_{\alpha}.
\end{equation} 
Thus, from the derivation of $l$ and $\tau_{\alpha}$ above
$
v_{if}\approx \vy{N}{H}\alpha \approx 600$~km~s$^{-1}$,
which is much faster than the speed of sound in the plasma.

The width of quasi-static IFs is of the order
of the mean free path for ionizing radiation, i.e. 
$1/\vy{n}{H}\vy{\sigma}{H}$, where $\vy{\sigma}{H}\approx 6\times 10^{-18}$~cm$^{2}$
is the photoionization cross section.  
Thus, the IF's in QSO~ 2359--1241 and SDSS~J0318--0600 are on the order of
$10^{13}$--$10^{14}$~cm, which are much smaller than the distance on which the IF retrats for flux variations on $\sim 1$~year time scale.
Still, these continuum variations
much shorter than the recombination time will broaden the IF creating an 
extended recombining region. But, as the IF departs from ionization equilibrium it
also departs from thermal equilibrium. In a static IF the temperature drops to a much lower value for a given value of the optical depth to  heating photons.
The optical depth is
\begin{equation}
\tau = \vy{\sigma}{H} \int_0^{S_{max}} \vy{n}{H^0} ds',
\end{equation}
with $\vy{n}{H^0}$ the 
neutral hydrogen density. 
Let us assume, for instance, a static IF of the form $\vy{n}{H^0}=\vy{n}{H}\times (1-e^{-s/\delta})$.  
As the ionized region shrinks the characteristic depth of the front increases to
$\delta'>\delta$.
At an early time $t$ after an instantaneous retreat of the IF 
$\vy{n}{H^0}\approx t/t_{\alpha}$. Thus, for an IF brodening by a factor 
$\nu$ it will take a time $t\approx t/\nu$ for the cooling front to start moving with the retreating IF.
Then, if the value of $\tau$ for which the temperature 
drops remains constant the depth within the IF at which this happens,
$S_{max}$, will be reduced with respect to $\delta'$. 
For example, if the change in temperature occurred at $S_{max}/\delta=1$ in the
static IF, $S_{max}/\delta'=0.68$ for $\delta'/\delta=2$, 
$S_{max}/\delta'=0.41$ for $\delta'/\delta=5$,
and $S_{max}/\delta'=0.28$ for $\delta'/\delta=10$. 
In the region of new lower equilibrium temperature cooling of the
plasma proceeds much faster than recombination,
because the cooling time scale is only
$\tau_T\approx 2\times10^4/\vy{n}{H}$~years \citep{spitzer}.
In conclusion,
the drop in temperature closely follows the supersonic shrinking of the ionized
region despite the broadening IF. 
Note that a drop in temperature makes recombination speed up, because 
$\alpha$ increases as $T_e^{-1/2}$ as the electron temperature 
$T_e$ decreases. 

What happens when the flux increases again? As the ionizing flux rises the 
neutral fragment of the cloud will be re-ionized. Thus, an IF will travel
through the neutral fragment, in the direction away from the ionizing source.
This process is governed by the equation
\begin{equation}
{\Delta Q_{H}\over 4\pi R^2} = \vy{n}{H^0} \vy{v}{IF}.
\end{equation}
The speed of the IF in this case is proportional by the change in radiative
flux and can be 
highly supersonic, in which case it will
send shocks ahead of the IF and rarefaction waves in the opposite direction.

In conclusion, time variations of the ionizing flux are expected to have two kinds of effects: 
(1) In the region neighboring the IF the ionization 
varies with time and departure from photoionization equilibrium could occur. This effect is out of the scope of this paper, but will be the subject 
of future investigation. 
(2) The varying ionizing flux will create supersonic cooling and heating fronts traveling through the ionized fraction of the cloud and the neutral region.
These will be
accompanied by rarefaction 
waves and shocks. 
This process is investigated in Section 4 of this letter, but
first 
we will derive a more accurate description 
of the dynamics of the IF.

\section{Dynamics of ionization fronts within outflowing clouds}

The IF within a cloud is defined as the region where the ionization drops to near zero. Thus, 
we solve for $n_e$ in the time dependent photoionization equation \begin{equation}
{d n_e\over d t} + \nabla . (n_e u) = \Gamma - \Lambda,
\end{equation}\label{continuity} 
Where $t$ is time, $u$ is the velocity of the gas, and $\Gamma$ and $\Lambda$ 
are the ionization and recombination rates per volume, 
\begin{equation}
\Gamma=-\nabla . \Phi_H=-\nabla .{\vy{Q}{H} e^{-\tau(x)}\over 4\pi R^2},
\end{equation}
\begin{equation}
\Lambda= \vy{\alpha}{H}  n_e^2
\end{equation}
where $\tau(x)$ is the optical depth within the cloud up to the depth $x$. 

Integrating Eqn.~\ref{continuity} in
the co-moving frame of the IF ($u=0$) over the ionized volume $V=\Omega R^2(l-r)$, with $\Omega$ the covering solid angle, 
one finds a differential equation for $r(t)$
\begin{equation}
{d(l-r(t))\over dt} = {\vy{Q}{H}\over 4\pi n_e R^2} - {l-r(t)\over \tau_\alpha}.
\end{equation}\label{ifeqn}
For gas clouds much smaller than the distance to the ionizing
source, i.e. $l\ll R$, one can assume constant density during a crossing
time of the cloud ($\sim l/v$). Then,
\begin{equation}
l-r(t)=l e^{-t/\tau_\alpha} \left[{1\over \tau_\alpha Q_H(0)} \int_0^t Q_H(t') e^{t'/\tau_\alpha}dt'+1\right].
\end{equation}\label{ifsolution}

To understand this solution it is useful to check a couple of cases:

\noindent{\it 1) $\vy{Q}{H}(t)$ varying as a step function.} 
Let us assume that at $t=0$ the absorber just reached $R_{if}$ 
when the continuum flux was $Q_0$. 
Thus, the cloud is
fully ionized and a drop in $Q_H$ will produce an IF. If the absorber
were closer than $R_{if}$ it could remain fully ionized even when $Q_H$
is reduced. 
If the flux drops to
$\xi Q_0$ at $t=0$ for $t>0$ one gets 
\begin{equation}
l-r(t)=l\times\left[\xi+(1-\xi) e^{-t/\tau_\alpha}\right]
\end{equation}
and 
\begin{equation}
v_{IF}=(1-\xi) {l\over \tau_\alpha}e^{-t/\tau_\alpha},
\end{equation}
which is consistent with the result of equation (4) and yields 
supersonic $v_{IF}$ for flux variations of $\sim 5\%$ or greater, i.e.
$\xi \le 0.95$.

\begin{figure}
\rotatebox{-90}{\resizebox{\hsize}{\hsize}
{\plotone{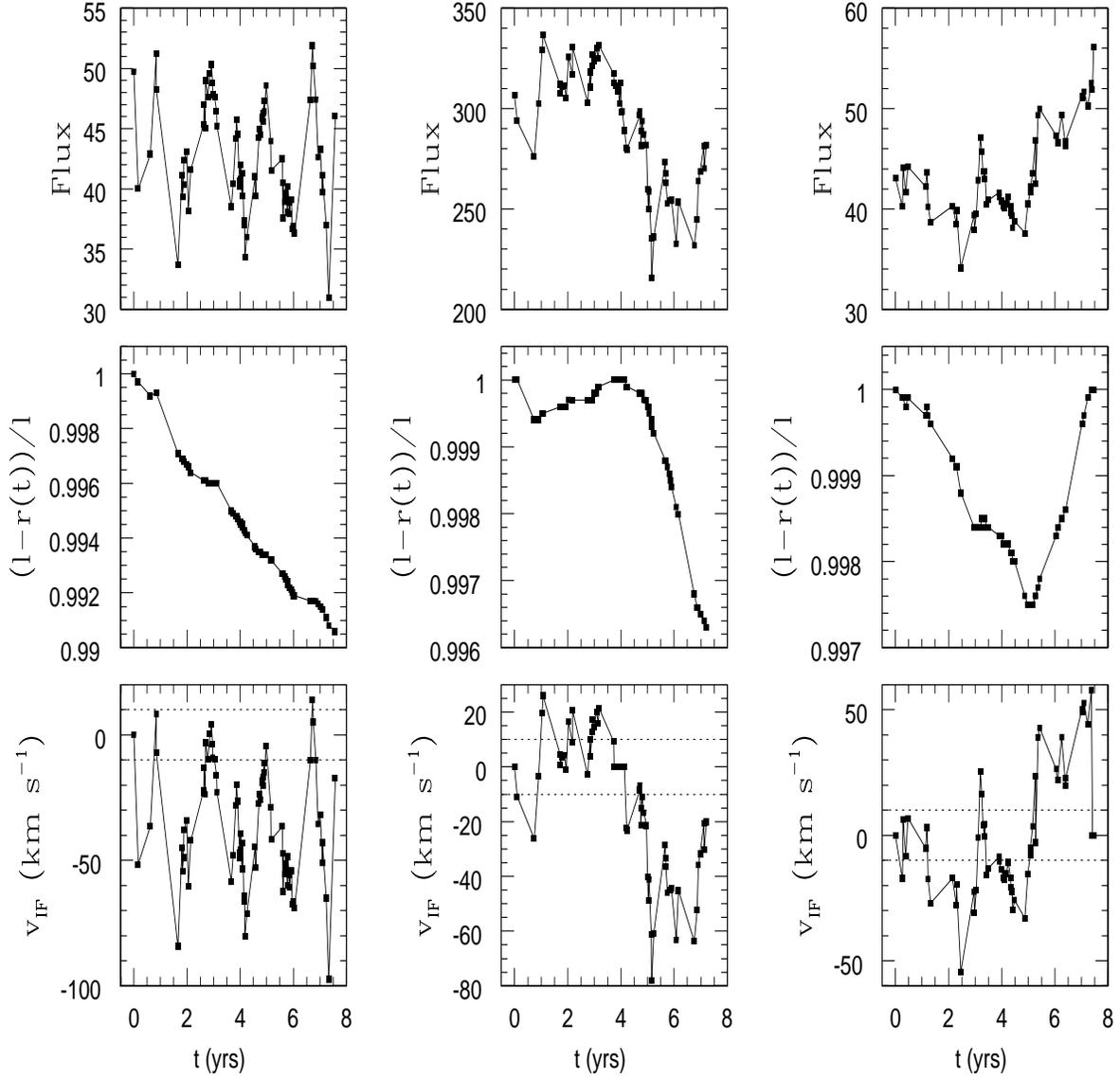}}}
\caption{Flux, $r(t)$, and $v_{IF}(t)$ for a sample of PG quasars
observed by \cite{kaspi}. The results are for PG0844 (left panels),
PG1226 (middle panels), and PG1411 (right panels). The calculations
assume $\tau_\alpha = 120$~yrs and $l=10^{17}$~cm. The horizontal dotted
lines in the $v_{IF}$ vs. $t$ plots delimit the approximate range for
subsonic motion.}
\label{ifobs}
\end{figure}

\noindent{\it 2) $Q_H(t)$ from observed quasar continua}. 
Here we compute the ionization of the cloud vs. time for selected 
quasar continua observed by \cite{kaspi} using Eqn.(10). 
Fig.~\ref{ifobs} shows the observer flux, the IF location with respect to
the total depth of the cloud $l$ and $v_{IF}$ 
for three quasars. These objects are chosen out of a
sample of 17 quasars observed by Kaspi et al. because these illustrate the
types of dynamical behaviors of the IF.
Because the calculations are done for
illustrative purposes only we 
disregard the uncertainties on the measured fluxes. 
Notice, for example, that in PG0844 the $r(t)/l$ ratio is generally decreasing with time
because the average ionizing continuum is lower than at the initial time
of observation. This overall behavior is somewhat arbitrary
as it depends on the initial observation. What is more meaningful, though,
is the instantaneous velocity of the IF front, 
computed as $\Delta r/\Delta t$, and this is seen to reach 
highly supersonic values. 
This is because the flux levels in PG0844, PG1226, and PG1411 vary by $\sim$40\%,
$\sim$50\%, and $\sim$30\% with respect to the value at $t=0$ in
less than a year. 
Clearly, this determination of velocities is sensitive to the observational
errors as well as the sampling of the quasar luminosity. Hence, we 
make no definitive claims on specific objects. However, the 
prediction that highly
supersonic fronts are expected is supported in every object of the available sample.

\section{Supersonic ionization fronts and cooling fronts}

The supersonic IF is accompanied 
by a cooling front 
that leaves a pressure step behind. This 
creates a rearward-facing, i.e. in the opposite direction to the IF, rarefaction wave and a forward-facing shock
\citep{strachan, ahlborn}.
Under homogeneous conditions and for an infinite flow stream 
these waves and shocks will form periodically \citep{schnerr}. 
As discussed before, the fragments once created become re-ionized and re-heated when the flux recovers. Then one can study the jump conditions between the 
main cloud and the fragments assuming constant temperature. 

In one dimension the jump conditions across the rarefaction wave are 
\begin{equation}
\rho_1 w_1 = \rho_2 w_2
\end{equation}\label{maseqn}
\begin{equation}
\rho_1 w_1^2 + P_1 +{B_1^2\over 8\pi} = \rho_2 w_2^2 + P_2 + 
{B_2^2\over 8\pi}
\end{equation}\label{preseqn}
\begin{equation}
B_1 w_1 = B_2 w_2
\end{equation}\label{mageqn}
where $\rho_i$ is the mass density of the gas, 
$w$ is the bulk velocity of the gas 
in 
the co-moving frame of the wave ($w=v-v_{wave}$), $P$ is the thermal pressure, 
and $B$ is the strength of the magnetic field. 
Because the jump is essentially isothermal we adopt $a^2=P_1/\rho_1=P_2/\rho_2$, 
and from Eqns. (13) - 
(15) one gets
\begin{equation}
w_1^2=\left({\rho_2\over \rho_1}\right)\left[a^2+ u^2_{A1}\left({\rho_2\over \rho_1}+1\right)\right],
\end{equation}
where 
 $u_{A1}=B_1/(8\pi \rho_1)^{1/2}$ is the Alfven velocity. 
Thus, the difference 
of velocities between the fragments and the main cloud is
\begin{equation}
v_2-v_1 = w_2-w_1 \approx 
\pm 20\times \left[1+{5\over 4}\left({v_{A1}\over 13~{\rm km~s}^{-1}}\right)^2\right]^{1/2},
\end{equation}\label{fragvelocity}
where we have used typical conditions $\rho_1/\rho_2=4$ and $a\approx 13$~km~s$^{-1}$.
This result shows that rarefaction waves, originated by supersonic IF, yield
lower density fragments that detach from the main cloud with relative
velocities of tens of km~s$^{-1}$ in absence of magnetic fields and much larger
velocities when these are present.  

There is evidence suggesting the presence of magnetic fields in
BALs.
Large trough widths, of the order of 100~km/s, are observed in FeLoBALS,
with the largest widths being generally associated with the main
absorber
(see Table 1 of Bautista et al. 2010).
The widths are much greater than expected from thermal 
broadening and seem to indicate supersonic turbulent velocities 
which could only be sustained by 
equally fast Alfven waves. 
The alternative scenario that observed trough widths could 
be mostly due to bulk velocity gradients along the line of sight is 
inconsistent with the fact that clouds have been able to travel 
$\sim 1 - 10$~kpc away from the AGN without diffusing into the 
interstellar medium. Moreover, it is 
difficult to understand how bulk velocity gradients could yield similar trough profiles for ions with different ionization stages
(but see Arav et al. 1999).

Strong magnetic fields in BAL clouds have been proposed in the past.
\cite{dekool95} pointed out that magnetic pressures in BAL clouds must
exceed thermal and radiative pressures to confine the clouds when accelerated
up to thousands of km~s$^{-1}$. Their models for acceleration of 
BAL clouds yield magnetic field strengths
of the order of 10~mG.
Further, \cite{furlanetto} argue that BAL quasar outflows could carry
a large fraction of the magnetic energy of the intergalactic medium in clusters
of galaxies.
The source of such magnetic fields could be either from the accretion disc
of the AGN \citep{dekool95}, or from dynamo-like effects
in the plasmas such as in 
the solar winds and supernova remnants  
\citep{giacalone,
kim, balsara}. 

We find that magnetic fields of the order of 10~mG, i.e. Alfven velocities of
a few hundred km~s$^{-1}$,  would explain 
the observed velocity separations between the main and the minor components
as observed in QSO~2359--1241 and SDSS~J0318--0600 (see Bautista et al. 2010). 

Another interesting aspect of Eqn.~(17) is that it predicts fragments that are faster than the main cloud as well as fragments that are slower than 
the main cloud, as seen in FeLoBALs. For example, since $w_2=4\times w_1$ if $w_2-w_1=+300$~km~s$^{-1}$ 
and $v_{wave}=1000$~km~s$^{-1}$, then $v_1=1100$~km~s$^{-1}$ and $v_2=1400$~km~s$^{-1}$. By contrast, if $w_2-w_1=-300$~km~s$^{-1}$ then $v_1=900$~km~s$^{-1}$ and $v_2=600$~km~s$^{-1}$.  The faster clouds, in the rest frame of the
quasar, are expected to
outrun the main cloud and remain undisturbed by it. The survival of the slower clouds 
depends on the geometry of the problem. In one dimensional geometry
the slow clouds 
would be swept up by the main cloud. 
In two and three dimensions 
it is possible that magnetic fields produce surface tension in the clouds 
capable of withstanding the fluid bulk pressure. 
The deceleration of the flow along the axis of travel of the main
cloud and its fragments yields a bow shock, where the supersonic gas 
of the main cloud flows smoothly around the small fragment
(see Shu 1992).

\section{Discussion and Conclusions}

We studied the effects of quasar time variable UV flux
on FeLoBAL clouds.  
In these systems absorption from \ion{Si}{2} and \ion{Fe}{2} comes 
from a  region just before an IF. This IF is formed within the absorbing cloud 
when the cloud crosses a distance $R_{if}$ from the radiation source.
At this distance, when the radiation flux drops the
ionized region of the cloud will retreat at supersonic speeds. 
A cooling front will follow the ionized region 
and the IF will also retreat and broaden. This creates a pressure imbalance 
across the IF on time scales much shorter than the sound crossing time.
Thus, the cooling front yields a rarefaction wave that travels into the low
temperature region. To conserve momentum across the cloud the large pressure
imbalance is compensated by an increased speed of the cold gas relative to the front. 
Now, when the flux recovers the neutral rarefied region is re-ionized and
re-heated by an IF moving away from the ionizing source. This IF can be
highly supersonic, driving a shock ahead of it and a rarefaction wave in the
opposite direction. The shocked gas forms new clouds and stabilize at a temperature of $\sim 10^4$~K, like the main cloud. These clouds 
are detached in velocity space from the main cloud
and there are rarefaction waves 
moving between the clouds. 

There are two fundamental parameters in determining the speed of the retreating IF.
One is the ratio of the depth of the cloud to the recombination rate, 
$l/\tau_\alpha$, which in turn is proportional to the column density of the 
cloud, $\vy{N}{H}$. Thus, high column density clouds can have faster 
IFs. The other parameter is the ratio of the variability time scale of the ionizing source
to the recombination time. Flux variations on scales of $\sim 1$~year are enough
for significant effects in FeLoBALs with particle densities of $\sim 10^4$~cm$^{-3}$, while variability on shorter time scales would affect
higher density clouds. When the flux drops, 
broadening of the IF slows the cooling front down such that marginally 
supersonic IF may not lead to shocks and waves, but faster IFs will. 

The thermal pressure imbalance across the 
IFs alone yields velocities to the fragments of the order of $\sim$20~km~s$^{-1}$ with respect to the main cloud. Much higher velocities can be attained if sizable magnetic pressures are considered. 
Moreover, in two objects for which complete measurements exists,
QSO~2359--1241 and SDSS~J0318--0600, the observed separation between kinematic components is well reproduced by magnetic field strengths of the order of $\sim 10$~mG.

Our model predicts that the detached 
fragments can be either faster or slower than the main cloud, 
as observed in FeLoBAL spectra.  
Survival of the slower fragments depends on the geometry of the problem. 
We propose that the slower fragments may permeate the main cloud if they are clumpy with a small filling factor and/or if the magnetic surface tension of the clouds is large enough to make the clouds flow around each other.

The physical effects presented here may 
not be exclusive to FeLoBALs, but could have implications in a 
variety of galactic and extragalactic phenomena. For example, in cataclysmic
variables, variable stars with circumstellar envelopes such as $\eta$ Carinae, and 
in the narrow line regions of AGN.

\begin{acknowledgements}
We like to thank Nahum Arav and Kirk Korista for fruitful discussions. 
We also thank the anonymous referee for important suggestions.
\end{acknowledgements}


\begin{thebibliography}{20} 
\expandafter\ifx\csname natexlab\endcsname\relax\def\natexlab#1{#1}\fi


\bibitem[{Ahlborn} and {Strachan}(1973)]{ahlborn}
Ahlborn, B. and Strachan, J.D. 1973, Can. J. Phys. 51, 1416

\bibitem[{{Arav} {et al.}(2005)}]{arav05}
Arav, N., Kaastra, J., Kriss, G.A., Korista, K.T., Gabel, J., Proga,
D.\ 2005, ApJ 620, 665

\bibitem[{{Arav} {et al.}(1999)}]{arav99}
Arav, N., Korista, K.T., de Kool, M., Junkkarinen, V.T., Begelman, M.C.
1999, ApJ 516, 27

\bibitem[{{Arav, Li, and Begelman} (1994)}]{arav94}
Arav, N., Li, Zhi-Yun, Begelman, M.C. 1994, ApJ 432, 62

\bibitem[{{Arav} {et al.}(2008)}]{arav08}
Arav, N., Moe, M., Costantini, E., Korista, K.T., Benn, C., Ellison,
S.\ 2008, ApJ 681, 954

\bibitem[{{Balsara} and {Kim} (2005)}]{balsara}
Balsara, D.S. and Kim, J. 2005, ApJ 634, 390


\bibitem[{{Bautista et al.}\ (2010)}]{bautista10}
Bautista, M.A., Dunn, J.P., Arav, N., Korista, K., Moe, M., Benn, C. 2010,
ApJ 713, 25   

\bibitem[{{Bautista} and {Pradhan}(1998)}]{baupra98}
Bautista, M.A.\ \& Pradhan, A.K.\ 1998, ApJ 492, 650

\bibitem[{{Blum} and {Pradhan} (1992)}]{blum}
Blum, R.D., Pradhan, A.K., 1992, ApJS, 80, 425

\bibitem[{{Brotherton et al.}(2001)}]{brotherton01}
Brotherton, M.S., Arav, N., Becker, R.H., Tran, H.D., Gregg, M.D., White,
R.L., Laurent-Muehleisen, S.A.\ \& Hack, W.\ 2001, ApJ, 546, 134


\bibitem[{{de~Kool} and {Begelman}(1995)}]{dekool95} 
de Kool, M. and Begelman, M.C. 1995, ApJ\ 455, 448

\bibitem[{{Dufton} and {Kingston} (1991)}]{dufton}
Dufton,P.L., Kingston,A.E., 1991, MNRAS, 248, 827

\bibitem[{{Dunn} {et al.}(2010)}]{dunn09}
Dunn, J., Bautista, M.A., Arav, N., Moe, M., Korista, K.T.,
Constantini, E. Benn, C., Ellison, S., Edmonds, D. 2010, ApJ 709, 611 

\bibitem[{{Dunn} {et al.}(2008)}]{dunn08}
Dunn, J., Crenshaw, D.M., Kraemer, S.B., Trippe, M.L. 2008 AJ 136, 136

\bibitem[{{Ferland} { et al.}(1998)}]{cloudy}
Ferland, G.J., Korista, K.T., Verner, D.A., Ferguson, J.W., Kingdon, J.B., Verner, E.M.\ 1998, PASP 110, 761

\bibitem[{{Foltz}{ et al.}(1987)}]{foltz}
Foltz, C.B., Weymann, R.J., Morris, S.L., Turnshek, D.A.\ 1987, ApJ 347, 450

\bibitem[{{Furlanetto and Loeb} (2001)}]{furlanetto}
Furlaneto, S.R. and Loeb, A. 2001, ApJ 556, 619

\bibitem[{{Gabel}{\ et al.}(2006)}]{gabel06}
Gabel, J. R., Arav, N., Kim, T.  2006, ApJ, 646, 742

\bibitem[{{Ganguly} and {Brotherton} (2008)}]{ganguly}
Ganguly, R. \& Brotherton, M. S. 2008, ApJ, 672, 102 

\bibitem[{{Giacalone} and {Jokipii} (2007)}]{giacalone}
Giacalone, J. and Jokipii, J.R. 2007, ApJ 663, L41

\bibitem[{{Hamann}{\ et al.}(2008)}]{hamann08} 
Hamann, F., Kaplan, K.F., Hidalgo, P.R., Prochaska, J.X.,
Herbert-Fort, S.\ 2008, MNRAS 391, L39

\bibitem[{{Inoue, Yanazaki, Inutsuka} (2009)}]{inoue}
Hamann, F., Kaplan, K.F., Hidalgo, P.R., Prochaska, J.X.,
Herbert-Fort, S.\ 2008, MNRAS 391, L39

\bibitem[{{Kaspi}{\ et al.}(2000)}]{kaspi}
Kaspi, S., Smith, P.S., Netzer, H., Maoz, D., Jannuzi, B.T.,
Giveon, U. 2000, ApJ 533, 631

\bibitem[{{Kim} and {Balsara} (2006)}]{kim}
Kim, J. and Balsara, D.S. 2006, Astron. Nachr. 327, 433

\bibitem[{{Knigge}{\ et al.}(2008)}]{knigge}
Knigge, C., Scarini, S., Goad, M.R., Cottis, C.E. 2008,
MNRAS 386, 1426

\bibitem[{{Korista}\ {et al.}(2008)}]{korista08}
Korista, K.T., Bautista, M.A., Arav, N., Moe, M., Constantini, E., 
Benn, C.\ 2008, ApJ 688, 108

\bibitem[{{Moe}{ et al.}(2009)}]{moe09}
Moe, M., Arav, N., Bautista, M.A., Korista, K.T.\ 2009
ApJ 706, 525          

\bibitem[{{Nussbaumer} (1977)}]{nussbaumer}
Nussbaumer H., 1977, A\&A 58, 291

\bibitem[{Schnerr} \& {Adam}(1997)]{schnerr}
Schnerr, G.H. and Adam, S. 1997, J. of Thermal Scisnce, 6, 171


\bibitem[{{Shu} (1992)}]{shu}
Shu, F.H. 1992, "The Physics of Astrophysics. Volume II: Gas Dynamics", 
University Science Books, Mill Valley CA, USA

\bibitem[{{Spitzer} (1998)}]{spitzer}
Spitzer, L. in Physical Processes in the Interstellar Medium, Ed. 
Wiley-Interscience Publication, New York, pp. 139

\bibitem[{Strachan} and {Ahlborn}(1975)]{strachan}
Strachan, J.D. and Ahlborn, B. 1975, Aust. J. Phys. 28, 395

\bibitem[{{Trump} (2006)}]{trump}
Trump, J. R. et al. 2006, ApJS, 165, 1

\bibitem[{{Turnshek}(1995)}]{turnshek}
Turnshek, D.A. 1995, in QSO Absorption Lines, ed. G. Meylan (Berlin:
Springer), 223



\bibitem[{{Voit}{\ et al.}(1993)}]{voit}
Voit, G.M. Weymann, R.J., Korista, K.T. 1993, ApJ 413, 95

\bibitem[{{Weymann}(1995)}]{weymann}
Weymann, R.J. 1995, in QSO Absorption Lines, ed. G. Meylan (Berlin:
Springer), 213

\bibitem[{{Wiese} and {Fuhr} (1995)}]{wiese}
Wiese W.L., Fuhr J.R., 1995, NIST Database
 for Atomic Spectroscopy, Version 1.0, NIST Standard Reference Database 61.


\end{thebibliography}
\end{document}